\documentclass{ws-rv9x6}

\usepackage{ws-rv-van}           

\begin{document}

\chapter{Electron and hole spin dynamics and decoherence in quantum
  dots}\label{sec4.5}

\author[D. Klauser et al.]{D. Klauser, D. V. Bulaev, W. A. Coish and Daniel Loss}

\address{Department of Physics and Astronomy, University of Basel,
  Klingelbergstrasse 82, CH-4056 Basel, Switzerland}

\begin{abstract}
In this article we review our work on the dynamics and decoherence of electron and hole
spins in single and double quantum dots. The first 
part, on electron spins, focuses on decoherence induced via the
hyperfine interaction while the second part covers decoherence and
relaxation of heavy-hole spins due to spin-orbit interaction as
well as the manipulation of heavy-hole spin using electric dipole spin
resonance. 
\end{abstract}

\section{Introduction}\label{subsec4.5.1}

The Loss-DiVincenzo proposal\cite{Loss:1998a} to use the spin of a  
single electron confined in a quantum dot as a qubit for quantum
computation has triggered significant interest in the dynamics and
control of single spins in quantum dots. This has led to numerous 
exciting experimental achievements, among them the realization of single
electrons in single dots \cite{Tarucha:1996a,Ciorga:2000a} as well
as double dots \cite{Elzerman:2003a,Hayashi:2003a,Petta:2004a}, the
implementation of single-spin read
out\cite{Elzerman:2004a,Hanson:2005a}, the demonstration of the
$\sqrt{\mathrm{SWAP}}$ operation via pulsed exchange
interaction\cite{Petta:2005b} and   the measurement of single-spin ESR
\cite{Koppens:2006a}. For a detailed account of the progress in
implementing the Loss-DiVincenzo proposal, see the extensive reviews in
Refs. \refcite{Cerletti:2005a} and \refcite{Coish:2006a}. 

On the theoretical side, one focus was, and still is, the investigation of the
decoherence induced by the nuclei in the host material via the
hyperfine interaction. The first part of this review article is devoted to
the discussion of the rich spin dynamics that results from the
hyperfine interaction. We first give an introduction to hyperfine
interaction in quantum dots (Sec. \ref{subsec4.5.2}). Subsequently, we
discuss dynamics under the influence of hyperfine 
interaction for the case of a single spin in a single
dot (Sec. \ref{subsec4.5.3}) and for a double dot with one electron in 
each dot (Sec. \ref{subsec4.5.4}). To conclude the part about
hyperfine interaction, we discuss the idea of narrowing the nuclear
spin state in order to increase the spin coherence time (Sec. \ref{subsec4.5.5}).

The second part of the article is devoted to the dynamics and the
manipulation of heavy-hole spins in quantum dots. The motivation to
study hole spins comes from the fact that the valence band has
$p$-symmetry and thus the hyperfine interaction with lattice nuclei
for holes is suppressed in comparison to that of the conduction band
electrons. As a consequence, the main interest for hole spin dynamics
is the relaxation and decoherence due to spin-orbit interaction and we
discuss this in Sec. \ref{subsec4.5.6}. The next task towards using
hole spins as qubits for quantum computation is of course the coherent
manipulation of single hole spins. A potentially powerful method to
achieve coherent manipulation of spins is electric dipole spin
resonance (EDSR). An analysis of EDSR for heavy holes in quantum dots
will be presented in Sec. \ref{subsec4.5.7}.

\section{Hyperfine interaction for electrons in quantum dots}\label{subsec4.5.2}

\begin{figure}
\begin{center}
\includegraphics[clip=true,width=3.5 cm,angle=270]{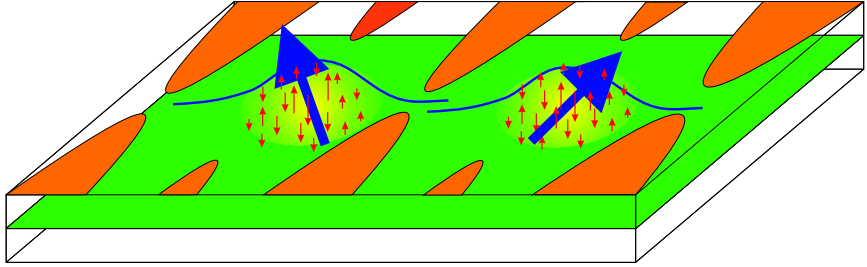}%
\caption{\label{fig:ddwn}A double quantum dot. Top-gates are set to a
  voltage configuration that confines the electrons in the
  two-dimensional electron gas (green) to quantum dots (yellow). The blue line
  indicates the envelope wave function of the electron (blue arrow). The hyperfine
  interaction with a particular nuclear spins (red arrows) is proportional to
  the envelope wave function squared at the position of the nuclei.
  Thus the nuclear spins in the center are drawn bigger since they
  couple stronger to the electron spin.}
\end{center}
\end{figure}

In this part of the article concerning electron spin decoherence we
assume that the orbital level spacing is much larger than the typical
energy scale of the hyperfine interaction. This is the case in typical
lateral quantum dots containing single electrons and allows one to
write an effective hyperfine Hamiltonian $H_{\mathrm{hf}}$ for a
single electron confined to the quantum-dot orbital ground state $\psi_0$
\begin{equation}
H_{\mathrm{hf}} = \mathbf{h}\cdot\mathbf{S}, \,\,\,\mathbf{h}=A\nu
\sum_k|\psi_0(\mathbf{r}_k)|^2\mathbf{I}_k,
\end{equation}
where $\mathbf{S}$ is the spin-1/2 operator for a single electron and
$\mathbf{I}_k$ is the spin operator for the nuclear spin at lattice
site $k$, while $\nu$ is the volume of the crystal unit cell and $A$
is the hyperfine coupling strength. For GaAs, which is mostly used for
the fabrication of lateral dots, the average hyperfine coupling strength
weighted by the natural abundance of each isotope is $A\approx 90\mu
eV$ \cite{Paget:1977a}. In Fig. \ref{fig:ddwn} the hyperfine coupling
of the electron spin in a lateral double quantum dot is illustrated. 
The electron spin dynamics under $H_{\mathrm{hf}}$ have been studied
under various approximations and in different parameter regimes. For
an extensive overview, see reviews in
Refs. \refcite{Cerletti:2005a}, \refcite{Schliemann:2003a} and \refcite{Taylor:2006a}. Here,
we briefly mention parts of this study before we focus on a few cases
of special interest. The first analysis of electron spin dynamics under
$H_{\mathrm{hf}}$ in this context showed that the long-time longitudinal spin-flip
probability is $\sim 1/p^2N$ \cite{Burkard:1999a}, i.e., this
probability is suppressed
in the limit of large nuclear spin polarization $p$ and large number $N$
of nuclear spins in the dot. An exact solution for the case of full
polarization ($p=1$) gives, for both transverse and longitudinal
electron spin components, a long-time power law decay $\sim 1/t^{3/2}$ by a
fraction $\sim 1/N$ on a timescale of $\tau\sim\hbar N/A\sim 1 \mu s$
(for a GaAs dot with $N \sim 10^5$)\cite{Khaetskii:2002a}. The fact that
this exact solution shows a non-exponential decay demonstrates the
non-Markovian behavior of the nuclear spin bath. For non-fully
polarized systems $p < 1$ and in the limit of large magnetic fields
(or high polarization $p\gg 1/\sqrt{N}$), the transverse electron spin
undergoes a Gaussian  decay\cite{Khaetskii:2002a,Merkulov:2002a} on a
timescale $\tau\sim \hbar\sqrt{N}/A\sqrt{1-p^2}$ ($\tau\sim 5 ns $
for GaAs with $p\ll 1$ and $N\sim 10^5$)\cite{Coish:2004a}. This Gaussian
decay can be reversed using a spin-echo sequence or by preparing the nuclear 
spin system in an eigenstate of $h_z$ \cite{Coish:2004a}. 

Several methods to prepare
the nuclear spin system have recently been suggested
\cite{Stepanenko:2006a,Klauser:2006a,Giedke:2006a} and we discuss
one of these methods \cite{Klauser:2006a} in
Sec. \ref{subsec4.5.5}. Once the nuclear spin system is prepared in an
eigenstate, the electron spin coherence is 
on one hand limited by dynamics in the nuclear spin system driven by
the dipole-dipole interaction for which a
worst case estimate \cite{Klauser:2006a} gives $\sim 100\mu s$ and on
the other hand, even for an eigenstate of $h_z$, there is decoherence
due to the flip-flop dynamics which can be important at times $\sim
\hbar N/A\sim 1\mu s$ (or less, depending on the size of the electron Zeeman
splitting). For the decay of nuclear spin polarization experiments
suggest timescales up to tens of seconds\cite{Koppens:2005a,
  Maletinsky:2007a,Baugh:2007a,Reilly:2007a} and hysteretic behavior of the 
nuclear spin polarization with respect to the external magnetic field
has been observed \cite{Maletinsky:2006a}. Further, measurements of the 
transport current in the so-called spin-blockade regime \cite{Ono:2002a} 
revealed hysteretic behavior with respect to the magnetic field \cite{Ono:2004a,Koppens:2005a} 
and bistable behavior in time\cite{Koppens:2005a}, which is attributed to a 
bistability in the nuclear spin polarization.
Further, very recent experiments suggest a
strong dependence of the nuclear field correlation time depending on whether
an electron is present in the dot or not and thus hyperfine mediated nuclear
spin flips are a possible mechanism for nuclear spin
diffusion\cite{Maletinsky:2007a}. This last mechanism has been
estimated to lead to fluctuations of nuclear spin polarization
on a timescale of $\sim 100 \mu s$\cite{Klauser:2006a}.

\section{Single-electron spin decoherence}\label{subsec4.5.3}
In this section we look in more detail at hyperfine-induced decoherence for a
single spin in a quantum dot in the regime of large Zeeman splitting
$\epsilon_z=g\mu_Bb_z$ (due to an externally applied magnetic
field $b_z$). If $\epsilon_z$ is much larger than
$\sigma=[\mathrm{Tr}\{\rho_I(h_z-h_0)^2\}]^{1/2}$,
with $h_0=\mathrm{Tr}\{\rho_Ih_z\}$ (where $\rho_I$ is the density
matrix of the nuclear spin system), we may neglect the transverse term
$S_{\bot}\cdot h_{\bot}$ and find that the Hamiltonian is simply
\begin{equation}
H_0=(\epsilon_z+h_z)S_z.
\end{equation}
This Hamiltonian just induces precession around the $z$-axis with a
frequency that is determined by the eigenvalue $h_z^n$ of $h_z$, where
$h_z|n\rangle=h_z^n|n\rangle$. For a large but
finite number of nuclear spins ($N\sim 10^5$ for lateral GaAs dots)
the eigenvalues $h_z^n$ are Gaussian distributed (due to the central
limit theorem) with mean $h_0$ and variance $\sigma\approx\hbar A/\sqrt{N}$
\cite{Coish:2004a}.     
Calculating the dynamics under $H_0$ (which is valid up to a
timescale of $\sim\epsilon_z/\sigma^2\sim 1\mu s$, where the transverse
terms become relevant) leads to a Gaussian decay of the transverse electron
spin state $|+\rangle=(|\uparrow\rangle+|\downarrow\rangle)/\sqrt{2}$
\cite{Coish:2004a}:
\begin{eqnarray}\label{eq:cpp}
C^0_{++}(t)&=&\frac{1}{\sqrt{2\pi}\sigma}\int^{\infty}_{-\infty}dh_z^n
e^{\left(-\frac{(h_z^n-h_0)^2}{2\sigma^2}\right)}|\langle
n|\otimes\langle+| e^{(-iH_0t)}|+\rangle\otimes|n\rangle|^2\nonumber\\
&=&\frac{1}{2}+\frac{1}{2}e^{\left(-\frac{t^2}{2\tau^2}\right)}
  \cos\left[(\epsilon_z+h_0)t\right];\,\,\,\,\,\tau=\frac{1}{\sigma}
  =\sqrt{\frac{N}{1-p^2}}\frac{2\hbar}{A}.     
\end{eqnarray}
Here again, $p$ denotes the polarization and for an unpolarized GaAs
quantum dot with $N \sim 10^5$ we find $\tau\sim 5 ns$. 
Applying an additional ac driving field with amplitude $b$ along the $x$-direction leads to
single-spin ESR. Assuming again that 
$\epsilon_z\gg\sigma$, we have the Hamiltonian 
\begin{equation}\label{eq:hesr}
H_{\mathrm{ESR}}=H_0+b\cos(\omega t)S_x.
\end{equation}
 In a rotating-wave approximation (which is valid for
 $(b/\epsilon_z)^2\ll1$) the decay of the driven Rabi oscillations is 
given by\cite{Koppens:2007a}
\begin{equation}
C^{\mathrm{ESR}}_{\uparrow\uparrow}(t)\sim 1-C+
\sqrt{\frac{b}{8\sigma^2t}}\cos\left(\frac{b}{2}t+\frac{\pi}{4}\right)
+\mathcal{O}\left(\frac{1}{t^{3/2}}\right),
\end{equation}
for $t\gtrsim \mathrm{max} \left(1/\sigma,1/b,b/2\sigma^2\right)$ and
$\epsilon_z+h_0-\omega=0$. Here, 
$C^{\mathrm{ESR}}_{\uparrow\uparrow}(t)$ is defined in the same way as
$C^0_{++}(t)$ in Eq. (\ref{eq:cpp}). The time-independent constant is
given by $C=\exp(b^2/8\sigma^2)\mathrm{erfc}(b/\sqrt{8}\sigma)\sqrt{2\pi}b/8\sigma$. 
The two interesting features of the decay are the slow ($\sim
1/\sqrt{t}$) power law and the universal phase shift of
$\pi/4$. The fact that the power law already becomes valid after a
short time $\tau\sim 15 ns$ (for $b\approx\sigma$) preserves the coherence over a long
time, which makes the Rabi oscillations visible even when the Rabi
period is much longer than the timescale $\tau \sim 15ns$ for
transverse spin decay. Both the universal phase shift and the
non-exponential decay have recently been observed in experiment
\cite{Koppens:2007a}. 
In order to take corrections due to the transverse terms
$S_{\bot}\cdot h_{\bot}$ into account, a more elaborate
calculation is required. The Hamiltonian with flip-flop terms (but
without a driving field) takes the form
\begin{equation}
H_{\mathrm{ff}}=H_0+\frac{1}{2}(S_+h_-+S_-h_+). 
\end{equation}
In Ref. \refcite{Coish:2004a} a systematic
calculation taking into account these so-called flip-flop terms was
performed using a generalized master equation, valid in the limit of
large magnetic field or large polarization. This calculation shows that
even for an eigenstate of $h_z$, for which the Gaussian decay in
Eq. (\ref{eq:cpp}) vanishes, the electron spin undergoes nontrivial
non-Markovian decay on a timescale $\hbar N/A\sim 10 \mu s$. 

Other calculations \cite{Yao:2005a,Shenvi:2005a,Deng:2006a} give
microsecond timescales for the electron spin decoherence due to electron-nuclear
spin flip-flops processes. The results in Ref. \refcite{Yao:2005a} suggest that also the 
decoherence due to dynamics in the nuclear spin system via
electron mediated nuclear dipole-dipole interaction is suppressed by a
spin echo and thus that the spin-echo decay time may be considerably
different from the (not ensemble averaged) free-induction decay. 

\section{Singlet-triplet decoherence in a double quantum 
  qot}\label{subsec4.5.4} 
We now move on to discuss hyperfine induced decoherence in a double
quantum dot. The effective Hamiltonian in the subspace
of one electron on each dot is best written in terms of the sum and
difference of electron spin and collective nuclear spin operators:
$\mathbf{S}=\mathbf{S}_1+\mathbf{S}_2,\delta\mathbf{S}=\mathbf{S}_1-\mathbf{S}_2$   
and $\mathbf{h}=\frac{1}{2}(\mathbf{h}_1+\mathbf{h}_2),
\delta\mathbf{h}=\frac{1}{2}(\mathbf{h}_1-\mathbf{h}_2)$:
\begin{equation}
H_{\mathrm{dd}}(t)=\epsilon_z S_z +\mathbf{h\cdot S}+
\delta\mathbf{h\cdot}\delta\mathbf{S} +\frac{J}{2}\mathbf{S\cdot S}-J.
\end{equation}
Here, $J$ is the Heisenberg exchange coupling between the two
electron spins. Similar to the single-dot case, we assume that the
Zeeman splitting is much larger than $\langle 
\delta \mathbf{h}\rangle_{\mathrm{rms}}$ and $\langle\mathbf{h}_i 
\rangle_{\mathrm{rms}}$, where $\langle\mathcal{O}\rangle_{\mathrm{rms}}
=[\mathrm{Tr}\{\rho_I (\mathcal{O}-\langle\mathcal{O}\rangle)^2\}]^{1/2}$ is the
root-mean-square expectation value of the operator $\mathcal{O}$
with respect to the nuclear spin state $\rho_I$. Under these
conditions the relevant spin Hamiltonian becomes block diagonal  with
blocks labeled by the total electron spin projection along the
magnetic field $S_z$. In the subspace of $S_z=0$ (singlet
$|S\rangle$, and triplet $|T_0\rangle$) the Hamiltonian can
be written as \cite{Coish:2005a,Klauser:2006a}  
\begin{equation}\label{Hamiltonian}
H_{\mathrm{sz0}}(t)=\frac{J}{2}\mathbf{S}\cdot\mathbf{S}+(\delta
h_z+\delta b_z) \delta S_z
\end{equation}
Here, $\delta b_z$ is the inhomogeneity of the externally
applied classical static magnetic field with $\delta b_z \ll
\epsilon_z$, while the nuclear difference field $\delta h_z$ is Gaussian
distributed, as was $h_z$ in the single dot case. A full account of the
rich pseudo-spin dynamics under 
$H_{\mathrm{sz0}}(t)$ can be
found in Refs. \refcite{Klauser:2006a} and \refcite{Coish:2005a}. Here we only discuss
the most prominent features for $C^{\mathrm{sz0}}_{SS}(t)$, which gives
the probability to find the singlet $|S\rangle$, if the system was initialized to
$|S\rangle$ at $t=0$. The parameters that determine the
dynamics are the exchange coupling $J$, the expectation value of the
total difference field $x_0=\delta b_z+\delta h_0$ and the width of
the difference field $\sigma_{\delta}$ (with $\delta h_0=
\langle\psi_{\mathrm{I}}|\delta h_z|\psi_{I}\rangle$ and
$\sigma_{\delta}=\langle\psi_{\mathrm{I}}|(\delta h_z-\delta
h_0)^2|\psi_{I}\rangle^{1/2}$). For the asymptotics one finds that
the singlet probability does not decay to zero, but goes to a finite,
parameter-dependent value \cite{Coish:2005a}. In the case of strong
exchange coupling $|J|\gg\mathrm{max}(|x_0|,\sigma_\delta)$ the singlet
only decays by a small fraction 
quadratic in $\sigma_{\delta}/J$ or $x_0/J$:
\begin{equation}
C^{\mathrm{sz0}}_{SS}(t\rightarrow \infty) \sim \left\{
\begin{array}{l}
1-2\left(\frac{\sigma_{\delta}}{J}\right)^2,
\,\,\,|J|\gg\sigma_{\delta}\gg|x_0|,\\
1-2\left(\frac{x_0}{J}\right)^2,
\,\,\,|J|\gg|x_0|\gg\sigma_{\delta}.
\end{array}\right.
\end{equation}
At short times $C^{\mathrm{sz0}}_{SS}(t)$ undergoes a Gaussian decay on a
timescale $\sqrt{J^2+4x_0^2}/4|x_0|\sigma_{\delta}$ while at long
times $t\gg|J|/4\sigma_\delta^2$ we have a power law decay  
\begin{equation}
C^{\mathrm{sz0}}_{SS}(t)\sim C^{\mathrm{sz0}}_{SS}(t\rightarrow\infty)
+e^{-\frac{x_0^2}{2\sigma_\delta^2}}
\frac{\cos(|J|t+\frac{3\pi}{4})}  
{4 \sigma_\delta \sqrt{|J|}\,t^{\frac{3}{2}}}.
\end{equation}
As in the case of single-spin ESR, we again have a power-law decay, now
with $1/t^{3/2}$ and a universal phase shift, in this case:
$3\pi/4$. Measurements\cite{Laird:2006a} of the correlator $C^{\mathrm{sz0}}_{SS}(t)$
confirmed the parameter dependence of the saturation value and were
consistent with the theoretical predictions concerning the decay. 
Using the same methods, one may also look at transverse correlators in
the $S_z=0$ subspace and find again power-law decays and a universal
phase shift, albeit, with different decay power and different value
of the universal phase shift\cite{Klauser:2006a}. Looking at the
short-time behavior of the transverse correlators also allows one to
analyze the fidelity of the $\sqrt{\mathrm{SWAP}}$ gate\cite{Klauser:2006a}.

\section{Nuclear spin state narrowing}\label{subsec4.5.5} 

 The idea to prepare the nuclear spin system in order to prolong the
 electron spin coherence was put forward in
 Ref. \refcite{Coish:2005a}. Specific methods for nuclear spin state 
 narrowing have been described in  Ref. \refcite{Klauser:2006a} in the
 context of a double dot with  oscillating exchange interaction, in
 Ref. \refcite{Giedke:2006a} for phase-estimation of a single (undriven)
 spin in a single dot and in an optical setup in  
Ref. \refcite{Stepanenko:2006a}. Here, we discuss narrowing for the case of
a driven single spin in a single dot, for which the details are very
similar to the treatment in Ref. \refcite{Klauser:2006a}. The general idea behind state
narrowing is that the evolution of the electron spin system depends on the value of the 
nuclear field since the effective Zeeman splitting is given by $\epsilon_z+h_z^n$. 
This leads to a nuclear field dependent resonance condition $\epsilon_z+h^n_z -\omega=0$ 
for ESR and thus measuring the evolution of the electron spin system determines $h_n^z$ and thus
the nuclear spin state. 

We start from the Hamiltonian for single-spin ESR as given in
Eq. (\ref{eq:hesr}). The electron spin is initialized to the
$|\uparrow\rangle$ state at time $t=0$ and evolves under $H_{\mathrm{esr}}$
  up to a measurement performed at time $t_m$. The probability to find
  $|\downarrow\rangle$  for a given eigenvalue $h^n_z$ of the nuclear
  field operator ($h_z|n\rangle=h_z^n|n\rangle$) is then given by
\begin{equation}
P^n_{\downarrow}(t)=\frac{1}{2}\frac{b^2}{b^2+4\delta_n^2}\left[1-\cos\left(\frac{t}{2}
  \sqrt{b^2+4 \delta_n^2}\right)\right]
\end{equation}
where $\delta_n=\epsilon_z+h_z^n-\omega$ and $b$ is the amplitude of
the driving field. As mentioned above, in equilibrium we have a
Gaussian distribution for the eigenvalues $h_z^n$, i.e., for the
diagonal elements of the nuclear spin density matrix $\rho_I(h_z^n,0)=\langle
n|\rho_I|n\rangle=\exp\left(-(h_z^n-h_0)^2/2\sigma^2\right)/\sqrt{2\pi}\sigma$.
Thus, averaged over the nuclear distribution we have the probability
$P_{\downarrow}(t)$ to find the state $|\downarrow\rangle$, i.e., 
$P_{\downarrow}(t)= \int dh^n_z\rho_I(h_z^n,0)P^n_{\downarrow}(t)$. 
After one measurement with outcome $|\downarrow\rangle$, we thus find for 
the diagonal of the nuclear spin density matrix \cite{Peres:1993a}
\begin{equation}
\rho_I(h_z^n,0)\stackrel{|\downarrow\rangle}{\longrightarrow} 
\rho_I^{(1,\downarrow)}(h^n_z,t_m)=\rho_I(h_z^n,0)
\frac{P^n_{\downarrow}(t_m)}{P_{\downarrow}(t_m)}.  
\end{equation}
Assuming now that the measurement is performed in
such a way that it gives the time averaged value (i.e., with a time
resolution less than $1/b$) we have for the probability
$P^n_{\downarrow}$ of measurement result $|\downarrow\rangle$ as a
function of the nuclear field eigenvalue $P^n_{\downarrow}=
\frac{1}{2}\frac{b^2}{b^2+4\delta_n^2}$. Thus, by performing a
measurement on the electron spin (with outcome $|\downarrow\rangle$),
the nuclear-spin density matrix is 
multiplied by a Lorentzian with width $b$ centered around the $h^n_z$ that satisfies the resonance condition
$\epsilon_z+h_z^n-\omega=0$. This results in a narrowed nuclear
spin distribution, and thus an extension of the electron spin
coherence, if $b<\sigma$.
In the case of measurement outcome $|\uparrow\rangle$ we find 
\begin{equation}
\rho_I(h_z^n,0)\stackrel{|\uparrow\rangle}{\longrightarrow} 
\rho_I^{(1,\uparrow)}(h^n_z,t_m)=\rho_I(h_z^n,0)
\frac{1-P^n_{\downarrow}(t_m)}{1-P_{\downarrow}(t_m)}, 
\end{equation}
i.e., the Gaussian nuclear spin distribution
is multiplied by one minus a Lorentzian, thus reducing the probability
for the nuclear field to have a value matching the resonance
condition $\epsilon_z+h_z^n-\omega=0$. Due to the slow dynamics of the nuclear spin system (see
discussion at the end of Sec. \ref{subsec4.5.2}), many
such measurements of the electron spin are possible (with
re-initialization of the electron spin between measurements). Under the
assumption of a static nuclear field during $M$ such initialization
and measurement cycles we find
\begin{equation}
\rho_I(h^n_z,0)\longrightarrow \rho^{(M,\alpha_{\downarrow})}(h^n_z)=\frac{1}{N}\rho_I(h^n_z,0)
(P_{\downarrow}^n)^{\alpha_{\downarrow}}(1-P_{\downarrow}^n)^{M-\alpha_{\downarrow}},
\end{equation}
where $\alpha_{\downarrow}$ is the number of times the measurement
outcome was $|\downarrow\rangle$. The simplest way to narrow is to
perform single measurements with $b\ll\sigma$. If 
the outcome is $|\downarrow\rangle$, narrowing has been achieved.
Otherwise, the nuclear system should be allowed to re-equilibrate before the
next measurement\cite{Klauser:2006b}. In order to achieve a systematic
narrowing, one can envision adapting the driving frequency (and thus
the resonance condition) depending on the outcome of the previous
measurements. Such an adaptive scheme is described in detail in
Refs. \refcite{Stepanenko:2006a} and \refcite{Klauser:2006a}. With
this we conclude the part on hyperfine-induced decoherence of electron
spins in quantum dots and move on to the heavy holes.

\section{Spin decoherence and relaxation for heavy holes}\label{subsec4.5.6}
Now we consider the spin coherence of heavy holes in quantum dots. The
contact hyperfine interaction between lattice nuclei and heavy-hole
spin is much weaker than that for electrons, since the valence band
has $p$ symmetry. Thus (neglecting $sp$ hybridization) only the weaker anisotropic hyperfine interaction is present. Therefore, the decoherence due to hyperfine
interaction is suppressed for heavy holes and in this section we focus
only on the spin decoherence due to spin-orbit interaction induced by
heavy-hole - phonon coupling. 

From the two-band Kane model, the Hamiltonian for the valence band of
III--V semiconductors is given by 
\begin{equation}
\label{eq:Hbulk} H_\mathrm{bulk}=H_\mathrm{LK}+\eta\mathbf{J}\cdot\mathbf{\Omega}+H_\mathrm{Z},
\end{equation}
where $H_\mathrm{LK}$ is the Luttinger-Kohn Hamiltonian
\cite{LK}. The second term is the Dresselhaus spin-orbit coupling (due
to bulk inversion asymmetry) for the valence band \cite{OO,BLPRL2005}, 
$\textbf{J}=(J_x,J_y,J_z)$ are $4\times4$ matrices corresponding to
spin $3/2$, $\Omega_z=P_z(P_x^2-P_y^2)$, and 
$\Omega_x$, $\Omega_y$ are given by cyclic permutations. The last term
in Eq.~(\ref{eq:Hbulk}) $H_\mathrm{Z}=-2\kappa\mu_B\mathbf{B\cdot
  J}-2q\mu_B\mathbf{B}\cdot\mathbf{\mathcal{J}}$ is the Zeeman term
for the valence band \cite{Luttinger} ($\kappa$ and 
$q$ are the Luttinger parameters \cite{Luttinger} and
$\mathbf{\mathcal{J}}=(J_x^3,J_y^3,J_z^3)$). 
\begin{figure}[t]
\begin{center}
\includegraphics[clip=true,width=10 cm]{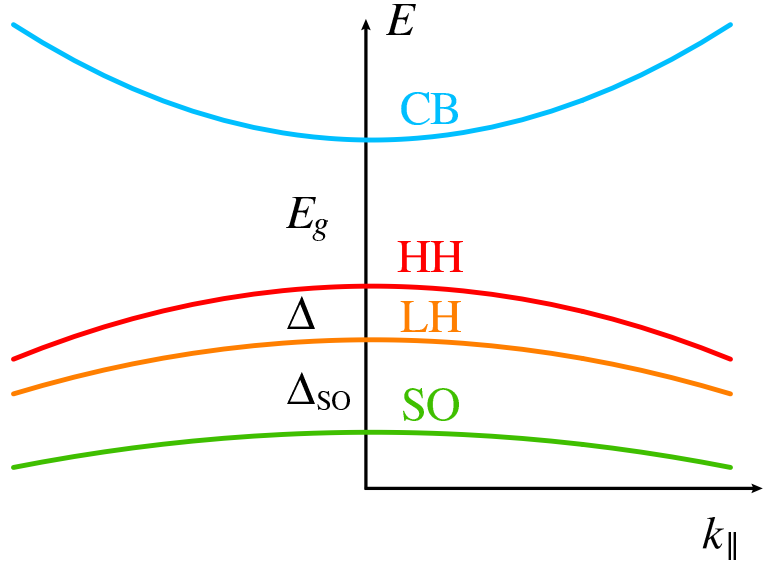}%
\caption{\label{fig:bandstr} Band structure of a III-V semiconductor quantum well with the $[001]$-growth direction, where $E_g$ is the band gap, $\Delta$ is the splitting between the light- and heavy-hole subbands due to quantum-well confinement, and $\Delta_\mathrm{SO}$ is the splitting of the valence band due to spin-orbit interaction.}
\end{center}
\end{figure}

We consider a $[001]$-grown two-dimensional system. In the case of an
asymmetric quantum well, due 
to structure inversion asymmetry along the growth direction, there is
an additional spin-orbit term, the Bychkov-Rashba spin-orbit term,
which, in the two-band model is given 
by \cite{Winkler2000} $\alpha_R\mathbf{P\times E\cdot J}$, where
$\alpha_R$ is the Bychkov-Rashba spin-orbit coupling constant and
$\mathbf{E}$ is an effective electric field along the growth
direction. 
Due to confinement along the growth direction, the valence
band splits into a heavy-hole subband with $J_z=\pm3/2$ and a
light-hole subband with $J_z=\pm1/2$ (see
Fig.~\ref{fig:bandstr} and Ref. \refcite{BLPRL2005}). If the
splitting $\Delta$ of heavy-hole and light-hole subbands is 
large, we describe the properties of heavy-holes and light-holes
separately, using only the $2\times2$ submatrices for the 
$J_{z}=\pm3/2$ and $J_{z}=\pm1/2$ states, respectively. The heavy-hole
submatrices have the property that 
$\tilde{J}_{x}=\tilde{J}_{y}=0$ and $\tilde{J}_{z}=\frac32\sigma_z$. For such a system and
low temperatures, only the lowest heavy-hole subband is significantly
occupied. In this case, we consider heavy holes 
only. In the framework of perturbation theory \cite{OO}, using
Eq.~(\ref{eq:Hbulk}) and taking into account the Zeeman 
energy and the Bychkov-Rashba spin-orbit coupling term, the
effective Hamiltonian for heavy holes of a quantum dot with lateral
confinement potential $U(x,y)$ is given 
by
\begin{equation}
\label{eq:H}
H=\frac{1}{2m}(P_x^2+P_y^2)+U(x,y)+H_\mathrm{SO}^\mathrm{HH}-\frac12g_{\perp}\mu_B
B_\perp\sigma_z, 
\end{equation}
where $m$ is the effective heavy-hole mass,
$\mathbf{P}=\mathbf{p}+|e|\mathbf{A(r)}/c$,
$\mathbf{A(r)}=(-yB_\perp/2,xB_\perp/2,yB_x-xB_y)$, $g_{\perp}$ is the
component of the $g$-factor tensor along the growth direction, and  
\begin{equation}
\label{eq:Hso} H_\mathrm{SO}^\mathrm{HH}=i\alpha P_-^3\sigma_++\beta
P_-P_+P_-\sigma_++\gamma B_-P_-^2\sigma_++\mathrm{H.c.} 
\end{equation}
is the spin-orbit coupling of heavy holes consisting of three
contributions:  the Dresselhaus term ($\beta$) \cite{BLPRL2005}, the
Rashba term ($\alpha$) \cite{Winkler2002}, and the last term
($\gamma$) combines two effects: orbital coupling via non-diagonal
elements in the Luttinger-Kohn Hamiltonian ($\propto P_\pm^2$) and
magnetic coupling via non-diagonal elements in the Zeeman term
($\propto B_\pm$). This latter term represents a new type of
spin-orbit interaction  which is unique for heavy holes
\cite{BLPRL07}. Here, $\alpha=3\gamma_0\alpha_R\langle
E_z\rangle/2m_0\Delta$, $\beta=-3\gamma_0\eta\langle 
P_z^2\rangle/2m_0\Delta$, $\gamma=3\gamma_0\kappa\mu_B/m_0\Delta$,
$\sigma_\pm=(\sigma_x\pm i\sigma_y)/2$, $P_\pm=P_x\pm iP_y$, $B_\pm=B_x\pm iB_y$, $m_0$ is
the free electron mass, $\gamma_0$ is the Luttinger parameter
\cite{Luttinger},  $\langle E_z\rangle$ is the 
averaged effective electric field along the growth direction of a
quantum dot, and $\Delta$ is the splitting of light-hole and
heavy-hole 
subbands. The splitting between heavy-hole and light-hole subbands
$\Delta\sim h^{-2}$, where $h$ is the quantum-dot height. 

\begin{figure}[t]
\begin{center}
\includegraphics[clip=true,width=10 cm]{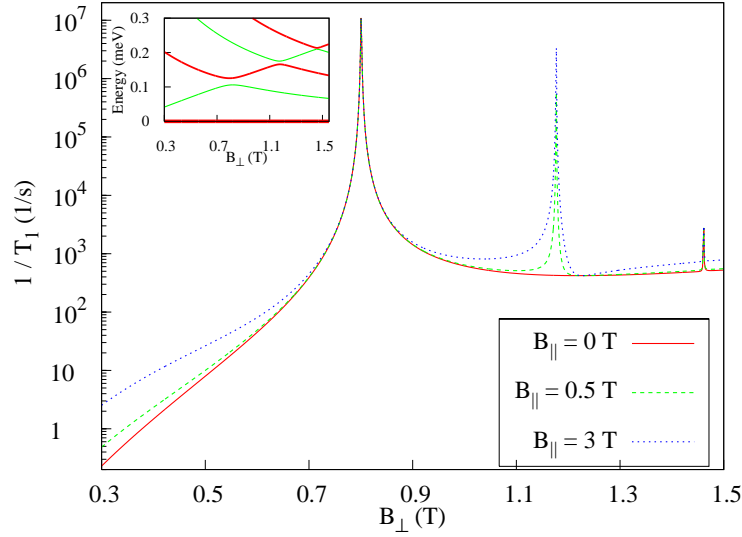}%
\caption{\label{fig:1}  Heavy hole spin relaxation rate $1/T_1$ in a
  GaAs quantum dot versus an applied perpendicular magnetic field
  $B_\perp$ (the height of the quantum dot is chosen to be $h=5\:$nm, the lateral
  size 
$l_0=\sqrt{\hbar/m\omega_0}=40\:$nm, $\kappa=1.2$, $\gamma_0=2.5$,
$g_\perp=2.5$ \cite{KCPF}, and the other parameters are given in
Ref.~\cite{BLPRL2005}). Inset: Energy differences of lowest excited
levels with respect to the ground state $E_{0,+3/2}$.  The second
avoided crossing comes from the spin-orbit interaction and the
in-plane magnetic field $B_\parallel$ (3rd term in
Eq.~\ref{eq:Hso}). The  anticrossing gap is proportional to
$B_\parallel$, implying that  the coupling between corresponding
states can be controlled 
externally.}
\end{center}
\end{figure}

The spectrum of (\ref{eq:H}) for parabolic lateral confinement $[U(x,y)=m\omega_0^2(x^2+y^2)/2]$ 
and for vanishing spin-orbit interaction
($H_{\mathrm{SO}}^\mathrm{HH}=0$) is the Fock-Darwin spectrum split
by the Zeeman term \cite{Fock,Darwin}. From Eq.~(\ref{eq:Hso}), it can
be seen that $H_{\mathrm{SO}}^\mathrm{HH}$ leads to coupling of the
two lowest states $|0,\pm3/2\rangle$ to the states with the opposite
spin orientations and different orbital momenta
$|l,\mp3/2\rangle$. Note that the three spin-orbit terms in
Eq.~(\ref{eq:Hso}) differ by symmetry in momentum space and hence mix
different states resulting in avoided crossings of the energy levels
(see inset of Fig.~\ref{fig:1}). Due to this spin-orbit mixing of the
heavy-hole states, the transitions between the states
$|0,\pm3/2\rangle$ with emission or absorption of an acoustic phonon
become possible and this is the main source of spin relaxation and
decoherence for heavy-holes \cite{BLPRL2005}. 

We consider a single-particle quantum dot, in which a heavy hole can
occupy one of the low-lying levels. In the following, we study the
relaxation of an $n$-level system, the first $n-1$ levels have the
same spin and the $n$-th 
level has the opposite spin orientation. In the framework of Bloch-Redfield theory \cite{Blum}, the Bloch equations for heavy-hole 
spin motion for such a system in the interaction representation are given by
\begin{eqnarray}
\label{eq:Sz}
\langle \dot{S}_z\rangle&=&\left(S_T-\langle S_z\rangle\right)/T_1-R(t),\\
\langle \dot{S}_x\rangle&=&-\langle S_x\rangle/T_2,\ \langle
\dot{S}_y\rangle=-\langle S_y\rangle/T_2, 
\end{eqnarray}
where $R(t)=W_{n1}\rho_{nn}(t)+\sum_{i=1}^{n-1}W_{ni}\rho_{ii}(t)$,
$\rho(t)$ is the density matrix, $W_{ij}$ is the 
transition rate from state $j$ to state $i$, $S_T$ is a constant
(which has the value of $\langle S_z\rangle$ in thermodynamic
equilibrium if $R(t)=0$),  
\begin{equation}
\label{eq:T1T2}
\frac{1}{T_1}=W_{n1}+\sum_{i=1}^{n-1}W_{in},\ \
\frac{1}{T_2}=\frac{1}{2T_1}+\frac12\sum_{i=2}^{n-1}W_{i1}, 
\end{equation}
where pure dephasing (due to fluctuations along $z$ direction) is
absent in the spin decoherence time $T_2$ since the spectral function
is superohmic. 
As can be seen from Eq.~(\ref{eq:Sz}), the spin motion has a complex
dependence on the density matrix and, in the general 
case, there are $n-1$ spin relaxation rates. However, in the case of
low temperatures ($\hbar qs_\alpha\gg k_\mathrm{B}T$), when  
phonon absorption becomes strongly suppressed, solving the master
equation, we find that $R(t)\approx0$. Therefore, there is only one spin relaxation time $T_1$. In this
limit, the last sum in Eq.~(\ref{eq:T1T2}) is 
negligible and the spin decoherence time saturates, i.e., $T_2=2T_1$.

 Note that in contrast to electrons \cite{GKL} there are no
 interference effects between different spin-orbit coupling terms,
 thus the total spin relaxation rate $1/T_1$ is the sum of rates
 $1/T_1=1/T_1^\mathrm{D}+1/T_1^\mathrm{BR}+1/T_1^{\parallel}$
 \cite{BLPRL2005,BLPRL07}: 
\begin{eqnarray}
\nonumber
\frac{1}{T_1^\mathrm{BR}}&\propto&\alpha^2\omega_Z^7\left(\frac{\omega_+^3}{3\omega_
++\omega_Z}-\frac{\omega_-^3}{3\omega_--\omega_Z}\right)^2\label{eq:WR},\\
\nonumber
\frac{1}{T_1^{\mathrm{D}}}&\propto&\beta^2\omega_Z^3\left(\frac{\omega_+}{\omega_
++\omega_Z}-\frac{\omega_-}{\omega_--\omega_Z}\right)^2\label{eq:WD},\\
\frac{1}{T_1^\parallel}&\propto&\gamma^2B_\parallel^2\omega_\mathrm{Z}^5\left(\frac{\omega_+^2}{2\omega_
++\omega_Z}+\frac{\omega_-^2}{2\omega_--\omega_Z}\right)^2, 
\label{eq:G1}
\end{eqnarray}
where $\omega_\pm=\sqrt{\omega_0^2+\omega_c^2/4}\pm\omega_c/2$,
$\omega_\mathrm{Z}=g_\perp \mu_BB_\perp/\hbar$,
$B_\parallel=\sqrt{B_x^2+B_y^2}$. 
In Fig.~\ref{fig:1} the total spin relaxation rate $1/T_1$ is plotted
as a function of perpendicular magnetic field $B_\perp$. There are
three peaks in the relaxation rate curve at
$\omega_\mathrm{Z}=\omega_-,\ 2\omega_-$, and $3\omega_-$, which are
caused by strong spin mixing at the anticrossing points. In the inset,
the first (third) avoided crossing resulting from Dresselhaus (Rashba)
spin-orbit coupling corresponds to the first (third) peak of the spin
relaxation curve in Fig.~\ref{fig:1}. At non-zero in-plane magnetic
fields ($B_\parallel$), there is an additional peak which is due to an
anticrossing between the energy levels $E_{0,+3/2}$ and $E_{2,-3/2}$
(see the second avoided crossing in the inset). Note that the spin
relaxation rate for heavy holes is comparable to that for electrons
\cite{BLPRL2005,BLPRB05} due to the fact that spin-orbit coupling of
heavy holes is strongly suppressed for flat quantum dots (see
Eq.~(\ref{eq:Hso})), as confirmed by a recent experiment \cite{Heiss2007}. 

\section{Electric dipole spin resonance  for heavy holes}\label{subsec4.5.7}

Let  us now consider methods for the  manipulation and detection of
the heavy-hole spin in quantum dots. For electrons in two-dimensional
structures, an applied oscillating in-plane magnetic field couples
spin-up and spin-down states via magnetic-dipole transitions and is
commonly used in electron spin resonance, Rabi oscillation, and spin
echo experiments \cite{Koppens:2006a}. It can be shown that
magnetic-dipole transitions ($\Delta n=0$, $\Delta \ell=0$, and
$\Delta s=\pm1$) are forbidden and, due to spin-orbit mixing of the
states $|0,\pm3/2\rangle$ with $i\beta_1^\pm|1,\mp3/2\rangle$,
electric-dipole transitions ($\Delta n=\pm1$, $\Delta \ell=\pm1$, and
$\Delta s=0$)  are most likely to occur. 
Therefore, the heavy holes are affected by the oscillating electric
field component and not by the magnetic one.  

We consider a circularly polarized electric field rotating in the
XY-plane with frequency $\omega$: $\mathbf{E}(t)=E(\sin\omega
t,-\cos\omega t,0)$. Therefore, the interaction of heavy holes with
the electric field is described by the Hamiltonian
$H^E(t)=(|e|E/m\omega)(\cos\omega tP_x+\sin\omega tP_y)$. The coupling
between the states $|\pm\rangle$ is given by
$\langle+|H^E(t)|-\rangle=H^E_{+-}=\left(
  H^E_{-+}\right)^*=d_{\mathrm{SO}}Ee^{-i\omega t}$, where  
\begin{equation}
d_{\mathrm{SO}}=(|e|l/2\omega)(\beta_1^-\omega_-+\beta_1^+\omega_+)
\label{eq:d_SO}
\end{equation}
is an effective dipole moment of a heavy hole depending on Dresselhaus
spin-orbit coupling constants, perpendicular magnetic field $B_\perp$,
lateral size of a quantum dot, and frequency $\omega$ of an rf electric field.

 In the framework of the Bloch-Redfield theory \cite{Blum} (taking
 into account also off-diagonal matrix elements), the effective master
 equation for the density matrix $\rho_{nm}$ assumes the form of Bloch
 equations \cite{Blum}, with the detuning of the rf field given by
 $\delta_{\mathrm{rf}}=\omega_\mathrm{Z}-\omega$. 
$2d_{\mathrm{SO}}E/\hbar$ is
the Larmor frequency, $T_1=1/(W_{+-}+W_{-+})$  the spin relaxation
time ($W_{nm}$ is the transition rate from state $m$ to state $n$),
$T_2=2T_1$ \cite{BLPRL2005} the spin decoherence time, and
$\rho_z^T=(W_{+-}-W_{-+})T_1$ the equilibrium value of $\rho_z$
without rf field. 

The coupling energy between a heavy hole and an oscillating electric
field is given by 
\begin{equation}
\langle H^E(t)\rangle= \mathrm{Tr}(\rho
H^E(t))=-\mathbf{d}_{\mathrm{SO}}\cdot\mathbf{E}(t), 
\label{eq:H^E}
\end{equation}
where
$\mathbf{d}_\mathrm{SO}=d_{\mathrm{SO}}(i\rho_{-+}-i\rho_{+-},\rho_{+-}+\rho_{-+},0)$
is the dipole moment of a heavy hole. Therefore, the rf power $P=
-d\langle H^E(t)\rangle/dt=-\omega d_{\mathrm{SO}}E\rho_-$ absorbed by
a heavy-hole spin system in the stationary state is given by
\cite{Abragam} 
\begin{equation}
P= \frac{2\omega
  (d_{\mathrm{SO}}E)^2T_2\rho_z^T/\hbar}{1+\delta_{\mathrm{rf}}^2T_2^2+(2d_{\mathrm{SO}}E/\hbar)^2T_1T_2}. 
\label{eq:P}
\end{equation}

\begin{figure}
\begin{center}
\includegraphics[clip=true,width=10 cm]{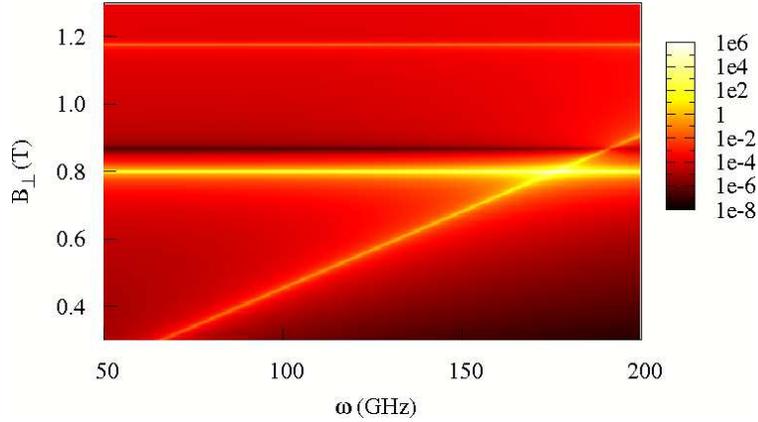}%
\caption{\label{fig:2} Absorbed power $P\:$(meV$/$s) as a function of
  perpendicular magnetic field $B_\perp$ and rf frequency $\omega$
  ($T_2=2T_1$, $E=2.5\:$V$/$cm, $B_\parallel=1\:$T, and the other
  parameters are the same as those in Fig.~\ref{fig:1}).} 
\end{center}
\end{figure}

In Fig.~\ref{fig:2}, the dependence of $P$ on a perpendicular magnetic
field $B_\perp$ and frequency $\omega$ of the oscillating electric
field is plotted. The rf power $P$ absorbed by the system has three
resonances and one resonant dip. The first resonance appears when the
energy of rf radiation equals the Zeeman energy of heavy holes:
$B_\perp^{\mathrm{r},1}=\hbar\omega/g_\perp\mu_\mathrm{B}$. The shape
of this resonance (at certain $\omega$) is given by 
$P\approx\hbar\omega\rho_z^T/2\hbar[1+\hbar^2\delta_{\mathrm{rf}}^2T_2/(2d_{\mathrm{SO}}E)^2T_1]$. 
 
If the first and second resonances are well separated
($\omega\ll\omega_-$), then the absorbed power can be estimated as 
\begin{equation}
P\approx 2\omega (d_{\mathrm{SO}}E)^2\rho_z^T/\hbar \delta_{\mathrm{rf}}^2T_2
\label{eq:P2}
\end{equation}
in the region of the second and third resonances and the resonant
dip. The second resonance corresponds to an anticrossing of the levels
$E_{0,-3/2}$ and $E_{1,3/2}$ (see the first avoided crossing in the inset
of Fig.~\ref{fig:1}) at $\omega_-=\omega_\mathrm{Z}$ \cite{BLPRL2005}
(at $B=B_\perp^{\mathrm{r},2}$). At the anticrossing point, there is
strong mixing of the spin-up and spin-down states and the dipole
moment of a heavy-hole spin system is maximal
$d_{\mathrm{SO}}^{\mathrm{max}}=|e|l\omega_\mathrm{Z}/2\omega$ and
equals half of the lowest electric dipole moment of a quantum dot
($|e|l\omega_-/\omega$). Therefore, the height of the second resonance
is given by
$(el\omega_\mathrm{Z}E)^2/2\hbar\omega\delta_{\mathrm{rf}}^2 T_2$. The
resonant dip appears at
$B_\perp^\mathrm{d}=(\hbar\omega_0/2g_\perp\mu_\mathrm{B})\sqrt{2m_0/g_\perp
  m}$, which corresponds to $\beta_1^-\omega_-+\beta_1^+\omega_+=0$
and to zero dipole moment (see Eq.~(\ref{eq:d_SO})). The third
resonance reflects the peak in the spin decoherence rate $T_2^{-1}$
due to an applied in-plane magnetic field (see Fig.~\ref{fig:1}) at the
second anticrossing point (the second avoided crossing in inset of
Fig.~\ref{fig:1}) at $2\omega_-=\omega_\mathrm{Z}$
($B_\perp^{\mathrm{r},3}=4\hbar\omega_0/g_\perp\mu_\mathrm{B}\sqrt{1+4m_0/g_\perp
  m}$). From the positions of the resonances we can determine
$g_\perp$, $m$, and $\omega_0$, from the shape and the height of those
we can extract information about the spin-orbit interaction constants
$\alpha$, $\beta$, and spin-orbit interaction strength due to in-plane
magnetic field (which is proportional to
$\gamma_0\kappa/\Delta$). Moreover, we can determine the dependence of
the spin relaxation and decoherence times on $B_\perp$.

\section{Conclusions}
We have discussed the rich dynamics of single electron spins in single and
double quantum dots due to hyperfine interaction with the nuclei. Key
features are non-exponential decays of various kinds and a remarkable universal
phase-shift. Further, we have studied spin decoherence and relaxation
of heavy holes in quantum dots due to spin-orbit coupling. The spin
relaxation time $T_1$ for heavy holes in flat quantum dots can be
comparable to that for electrons \cite{BLPRL2005,BLPRB05} as confirmed by experiment \cite{Heiss2007}. The spin
decoherence time for heavy holes is given by $T_2=2T_1$ at low temperatures. There is strong spin
mixing at energy-level crossings resulting in a non-monotonic dependence
$T_1(B)$. We have proposed a new method for manipulation of a
heavy-hole spin in a quantum dot via rf electric fields. This method
can be used for detection of heavy-hole spin resonance signals, for spin manipulation, and for determining important parameters of
heavy holes \cite{BLPRL07}.  

\bf{Aknowledgments:} \rm We aknowledge financial support from the Swiss
NSF, the NCCR Nanoscience and JST ICORP.

\bibliographystyle{ws-rv-van}

\bibliography{loss}

\end{document}